# The effect of pressure in the crystal and magnetic structure of FeWO$_4$

Oscar Fabelo[1], Javier Gonzalez-Platas[2], Stanislav Savvin[1], Pablo Botella[3], and Daniel Errandonea[3]

[1]Institut Laue-Langevin, 71 avenue des Martyrs, CS 20156, 38042 Grenoble cedex 9, France

[2]Departamento de Física, Instituto Universitario de estudios Avanzados en Física Atómica, Molecular y Fotónica (IUDEA), MALTA Consolider Team, Universidad de La Laguna, La Laguna, Tenerife, Spain

[3]Departamento de Física Aplicada-ICMUV, MALTA Consolider Team, Universitat de Valencia, Dr Moliner 50, Burjassot, Valencia, Spain

**Abstract:** The temperature dependence of the structural and magnetic properties of wolframite-type FeWO$_4$ were studied in situ by high pressure neutron diffraction. Neutron diffraction measurements were performed at the XtremeD instrument at the Institut Laue Langevin up to a maximum pressure of 8.7(4) GPa and a minimum temperature of 30.0(5) K. The diffraction data were analyzed via Rietveld refinements. We found that despite of producing a contraction of 5% of the volume, the maximum pressure applied in this study does not modify the Shubnikov space group below magnetic order. However, the orientation of magnetic moments and the Néel temperature, are slightly modified with the pressure, which is expected according to the preexistent understanding of magnetism in wolframites. We also determined a pressure-volume equation of state of FeWO$_4$ at 300 K, which is compared with previous X-ray diffraction studies and density-functional theory calculations.

# INTRODUCTION

The wolframite-type $MWO_4$ compounds form an interesting class of bimetallic oxides due to their properties such as magnetic for M = Fe, Co, and Ni and even multiferroic in the case of $MnWO_4$ [1-2]. Their magnetism is governed by the partially filled 3d orbitals of the divalent M cations [3]. In addition, to their interesting magnetic properties wolframites like $FeWO_4$, have been used to develop supercapacitors, photocatalytic, and photoluminescent materials [4] or to use magnetic and electric order in data store density applications in memory and optical devices [5-6]. In this sense, a deeper understanding of such properties is needed, where high-pressure studies may play a fundamental role, and $FeWO_4$ could be a good candidate for it.

The crystal structure of $FeWO_4$ is monoclinic at room temperature (RT) being described by space group *P2/c*, see Figure 1(a), and transforms to $P_a2/c$ below the magnetic order temperature [6,7]. It can be described as a slightly deformed hexagonal closest-packed lattice of oxygen atoms, with the metal ions ($Fe^{2+}$ and $W^{6+}$) occupying half of the octahedral holes. The crystal structure is represented in Figure 1(a). It can be described as two zigzag chains of edge-sharing $FeO_6$ or $WO_6$ octahedral units running along the [001] direction. The high-spin $d^6$ electronic configuration of the $Fe^{2+}$ cation distorts the $FeO_6$ octahedron by the Jahn–Teller effect. Magnetic measurements at ambient pressure revealed an antiferromagnetic Néel temperature ($T_N$) of *ca.* 75 K at ambient pressure and anisotropic properties related with the magneto-crystalline anisotropy of the $Fe^{2+}$ ion [6]. The magnetism of $FeWO_4$ is related to the presence of super-exchange interactions via one or two intervening oxygen ions along different paths between Fe…Fe neighbors [8]. The sign and magnitude of the interaction are determined by the participating magnetic ions ($Fe^{2+}$) and by the different magnetic exchange interactions [8].

It is a common knowledge that pressure could significantly alter the inter-atomic distances within solids, resulting in a range of intriguing phenomena [9]. Specifically, several studies have demonstrated that pressure can tune the magnetic order [10]. Over the past ten years, wolframites have been the subject of studies involving compression [11]. Because of their bulk modulus, a change up to 10% in the unit-cell volume is produced by the application of an external pressure of 10GPa. Consequently, the electronic, magnetic, vibrational, and elastic properties can be dramatically altered by pressures relatively small pressures compared to pressure achieved today in current high-pressure (HP) studies [12,13]. However, it is known that wolframite-type compounds do not collapse in a new structure at room temperature up to pressures of around 20 GPa [10-12]. Indeed, recent powder XRD experiments, performed at room temperature, have shown that $FeWO_4$ remains stable in the wolframite structure up to 20 GPa [14]. However, the linear compressibility changes, as a function of the applied pressure, in an anisotropic form. The structural changes induced by pressure cause a decrease of the electronic band-gap energy which has been attributed to the enhancement of hybridization between 3d electrons of the divalent M cation and 2p electrons of oxygen atoms [5, 11, 12]. Another interesting effect of pressure in wolframites are the changes induced by pressure in the $MO_6$ octahedra which affect the Jahn-Teller distortion [15], although this effect is expected to be weak in high spin $Fe^{2+}$ complexes.

The compressibility and anisotropic mechanical properties of wolframites, make $FeWO_4$ a quite interesting candidate to study the effect of pressure on the magnetic properties. According to density-functional theory (DFT) calculations, the magnetic properties of wolframites, particularly $FeWO_4$, should be also affected by pressure [16]. This is because super-exchange interactions between the $Fe^{2+}$ cations are expected to be modified due to changes induced by pressure in the $Fe^{2+}$ lattice [17]. However, the influence of pressure in the magnetism of wolframites has not been studied yet. Thus, $FeWO_4$ is an ideal candidate to perform a combined study of the influence of pressure on the crystallographic and magnetic structure.

$FeWO_4$ is our choice of study not only for the above-described reasons but also because the antiferromagnetic order of $FeWO_4$ is less complex than the magnetic order of related compounds like $MnWO_4$ [18]. In the antiferromagnetic configuration of $FeWO_4$, known as $AF_1$, see Figure 1, there are Fe atoms with parallel spins forming zigzag chains running along the c-axis. The zigzag chains are ferromagnetically coupled along the *b*-axis, forming ferromagnetic planes that are antiferromagnetically coupled along the *a*-axis. On one side, pressure can modify the Néel temperature ($T_N$), on the other it could modify the magnetic moment strength and orientation. Pressure can also transform the magnetic structure of $FeWO_4$ into a different antiferromagnetic order, like that observed in $MnWO_4$ with temperature [18]. This system exhibits an incommensurate spin-density wave state below 13.5K, which evolves into the so-called $AF_2$ phase below 12.3K. The wavevector hardly changes, but the presence of a spin component along the *b*-axis gives rise to an elliptical spiral structure. DFT calculations estimate that the energy for FM interactions in $FeWO_4$ is just slightly higher than AFM (*ca.* 10 meV per formula unit) [16]. Therefore, it would not be surprising that pressure could induce a change in some of the exchange couplings and consequently inducing magnetic transitions as predicted in the DFT calculations. Here we report a neutron diffraction study of $FeWO_4$ under high pressure up to 8.7(4) GPa and at low temperatures up to 30.0(5)K to investigate the influence of external pressure on the crystal and magnetic structure.

## EXPERIMENTAL DETAILS

Materials used as reagents in the synthesis of $FeWO_4$ powder were $Fe(NH_4)_2(SO_4)_2 \cdot 6H_2O$ (Aldrich, 99.0%), and $(NH_4)W_{12}O_{39}$ (Aldrich, 99.0%). The starting solutions containing $Fe^{2+}$ and $W^{6+}$ were prepared by dissolving their respective salts in distilled water. Then, they were combined to obtain a Fe-W source solution having a total cationic concentration of 0.3M, a total volume of 21.7mL. The masses of the different reagents were adjusted to obtain 1 g of the final product. Droplets of the obtained solution were flash frozen by projection onto liquid nitrogen and then freeze-dried at a pressure of 1-10 Pa and at a temperature of 228K in a Telstar Cryodos freeze-dryer. By following this method, a loose powder in the form of an amorphous solid precursor was successfully obtained. The thermal evolution of the precursor was monitored by means of thermogravimetric experiments under an oxygen atmosphere (heating rate 10 K min$^{-1}$, flow rate 60 cm$^3$

min$^{-1}$), carried out using a TG550 thermal analyzed from TA Instruments. The final FeWO$_4$ powder was obtained decomposing the precursor on an oven at 930ºC for 1 hour. The final product crystallizes in the *P*2/*c* space group with similar cell parameters than those previously reported by Ulku *et al.* [19], confirmed by powder X-ray diffraction. The structural model used in our refinements corresponds to a well-ordered wolframite structure. In the results presented by Maignan *et al.* [6], the authors describe the structure of FeWO$_4$ with two additional positions, W$_2$ and Fe$_2$. However, this cationic disorder was estimated to be approximately 1%, based on high-resolution X-ray data. In our case, we used high-flux instruments instead of than high-resolution ones, this level of disorder falls within the experimental error. Consequently, the refinements were carried out using the wolframite structure without accounting for cationic disorder.

The HP neutron measurements were performed on the XtremeD diffractometer at the Institut Laue Langevin (ILL). XtremeD was operated with the focusing pyrolytic graphite monochromator that provides neutrons of wavelength 2.45 Å. The large solid angle detector covers 130 degrees in 2θ and was equipped with an in-house designed radial oscillating collimator (ROC) that allows to reduce significantly the background coming from the environment of the sample. The HP conditions were obtained by using the ILL VX-5 Type Paris-Edinburg pressure cell reference 18PG1000BN5 [20]. The gaskets of the cell were made of the TiZr null matrix alloy (52.2% Ti / 47.8% Zr), which is transparent to neutrons. The dimensions of the gasket were Ø4.8x2.7mm, allowing us to increase the nominal pressure to nearly 10 GPa. Cubic boron nitride (cBN) anvils were used, and a mixture of deuterated ethanol-methanol was used as pressure-transmitting medium (PTM). The PTM frozen both at low-temperature and high-pressure generated non-hydrostatic stresses. However, we found that the broad of peaks is not affected by them which indicates that the collected data are of enough quality for the purposes of the present study. A small quantity of lead was used as a sensor for determining the pressure during experiments [21]. The cell was placed in a self-designed displex refrigerator (247ILCCHP180), which allows cool down the HP cell to 30.0(5) K. Temperature and pressure can be controlled by the instrument control program NOMAD. Due to the size of the gasket and the anvil, the maximum pressure safely reached to avoid technical issues was 8.7(4) GPa. However, in the first increase of pressure, the detector was partially shadowed in the upper and lower regions. These dark areas became more pronounced as the pressure increased, suggesting that the gaskets displaced the cadmium shielding of the cells. Although we attempted to minimize the impact of these shadows by selecting only the central part of the 2D detector, we are not confident that the relative intensities were correctly integrated. Nevertheless, these shadows do not affect the positions of the reflections, so the evolution of the cell parameters and the equation of state derived from these measurements should not be impacted by this artifact. Throughout this manuscript, quantitative analyses requiring precise measured intensity values will be conducted using the data obtained at ambient pressure and from the second load at 7.7(4) GPa.

Once the sample has been placed inside the pressure cell, a reference pressure was established through the equation of state (EoS) of lead according to the results reported by Strässle *et al*. [21], and subsequently the entire system was cooled, and spectra were taken for a period of 3 to 6 hours, depending on the pressure, using a 2theta-scan to improve the data quality. The decrease of pressure due to the temperature change from RT to 30 K was estimated to be in the order of 0.6 GPa. In this experiment, two cells were loaded, pressure changes were always performed at RT to avoid capillary blockage and ensure change of pressures under hydrostatic conditions. Data integration was done using Int3D software [22] and data analysis was done using FullProf [23].

## RESULTS AND DISCUSSION

To evaluate the crystal structure of the sample, a diffraction pattern was collected on the D1B instrument under ambient conditions. A Rietveld refinement was carried out to determine the cell parameters and the unit-cell volume. Results are shown in Figure 2. We confirmed the crystal structure corresponds to the expected wolframite-type structure (space group *P*2/*c*) with unit-cell paramters *a* =

4.7254(3) Å, $b$ = 5.6956(5) Å, $c$ = 4.9499(3) Å, $\beta$ = 89.792(10)º, $V$ =133.221(16) Å$^3$, in agreement with the literature [7, 14, 24].

New reflections are observed below $T_N$. These reflections have been indexed using the K-search program included in the FullProf suite [24]. The most likely propagation vector is **k** = (1/2, 0, 0), which correctly fits all magnetic reflections (see Figure 3). This means that the magnetic unit cell is two times the paramagnetic unit cell along the $a$-axis. To determine the possible magnetic structures compatible with the symmetry of the parent crystal structure of FeWO$_4$, described in the monoclinic $P12_1/c1$ space group and the **k** = (1/2, 0, 0) propagation vector, we have used the MAXMAGN utility of the Bilbao Crystallographic Server [25].

There are four maximal Shubnikov space groups, all of which give antiferromagneitic structures due to the application of the anti-translation operator along the $a$-axis. Each of these models corresponds to an irreducible representation, from $\Gamma_1$ to $\Gamma_4$, describing four different Shubnikov magentic space groups. All of them of the type $P_a2/c$. Of the four models, two allow magnetic moments in the $ac$-plane and in the other two the magnetic moments are constrained by symmetry to be along the $b$-axis. Therefore, the two Shubnikov groups ($\Gamma_1$ and $\Gamma_2$) have two degrees of freedom for magnetic moments (u, 0, w), while the other two ($\Gamma_3$ and $\Gamma_4$) have only one (0, v, 0). Table 1 presents a schematic representation of the four potential magnetic models, along with the corresponding transformation matrices. The orientation of the magnetic moments in each model is determined based on the symmetry constraints of each magnetic space group. For didactic purposes, arbitrary values for the magnetic moments have also been included in the models. It deserves to be noted that in all four models, there is only one iron orbit, and the application of the symmetry operators of each magnetic space group gives rise to the four iron positions described within the magnetic unit cell.

The four solutions have been used to refine magnetic reflections, but only $\Gamma_2$ is able to reproduce ther magnetic intensities. The results of the refinement using $\Gamma_2$ model is shown in Figure 4 form ambient pressure and for 7.7GPa. $\Gamma_2$ produces a magnetic structure that can be defined as ferromagnetic chains running along the

*c*-axis that are antiferromagnetically coupled with the adjacent chains along the *a*-axis. This magnetic structure is shown in Figure 5.

**Table 1:** Summary of the different magnetic models and transformation matrices

| | | |
|---|---|---|
| $\Gamma_1$ | | $\begin{pmatrix} 2 & 0 & 0 & 0 \\ 0 & 1 & 0 & 0 \\ 0 & 0 & 1 & 0 \end{pmatrix}$ |
| $\Gamma_2$ | | $\begin{pmatrix} 2 & 0 & 2 & \frac{1}{2} \\ 0 & 1 & 0 & 0 \\ -2 & 0 & -1 & 0 \end{pmatrix}$ |
| $\Gamma_3$ | | $\begin{pmatrix} 2 & 0 & 2 & 0 \\ 0 & 1 & 0 & 0 \\ -2 & 0 & -1 & 0 \end{pmatrix}$ |
| $\Gamma_4$ | | $\begin{pmatrix} 2 & 0 & 0 & \frac{1}{2} \\ 0 & 1 & 0 & 0 \\ 0 & 0 & 1 & 0 \end{pmatrix}$ |

We would like to indicate that the values found for the magnetic moment differ notably from those published by Ulku *et al.* [19], with **m** = (-1.92, 0.0, 1.06) resulting in a value of 2.19 μB, which is approximately half of the value observed in our fit (4.75(3) μB). However, the relative orientation of the magnetic moments is in good agreement in both refinements, with the component along the *a*-axis being almost twice as large as the component along the *c*-axis. The origin of the difference in the value of the magnetic moment is not clear but it is likely related to an inadequate determination of the scaling factor between the nuclear and magnetic parts in the work of Ulku *et al*. [19]. It should be noted here that our magnetic moment agrees also with the vale obtained by Maignan *et al*. [6] from susceptibility measurements (5.2 μB).

The refined magnetic moment at 7.7GPa was fitted in the critical region ($|T-T_N|/T_N$<0.1) using a power law $M \propto |T - T_N|^\beta$ (Figure 6) giving rise to a critical exponent $\beta$ of 0.43(4). This number should be taken with care, due to the low number of experimental points to fit in the critical region, the obtained value is slightly higher than expected value for a 3D Heisenberg model (β=0.365) but below of the expected for a mean-field.

Additional support to out refined value of the magnetic moment (4.75(3) μB) comes from the fact that it is very similar to the value expected for an S = 2 ion with spin-only contribution (4.0 μB). In the case of $Fe^{2+}$, the angular momentum is not zero, so it can present a relatively strong spin-orbit coupling, which causes a magnetic anisotropy and a non-quenched angular momentum that influences the total magnetic moment. The orbital magnetic contribution can be influenced by a Jahn-Teller distortion, which is quite common in $Fe^{2+}$ compounds. In $FeWO_4$, the octahedral environment of Fe is distorted, presenting 4 long distances (Fe-O2) = 2.191(18) Å and two short distances (Fe-O1) = 2.04(3) Å. Thus, the $FeO_6$ octahedron presents a moderate distortion with a Fe-O mean distance ($d_{mean}$) of 2.12(3) Å, a distance distortion (ζ) of 0.14(6) Å, and an angle distortion (Σ) of 61.4(5)º, defined as the summation of the different between the ideal value (90º) and each of the twelve possible cis O-Fe-O angles. In addition, the octahedron has a torsional distortion (Θ) of 128(2)º. This value is calculated as the summation by pairs of the ideal value (60º) minus the twenty-four possible dihedral angles between the atom vectors defined by two twisting triangular faces [26]. Therefore, this distortion may be responsible for the difference between the values obtained for the magnetic moment and those expected for a high spin $Fe^{2+}$ (S = 2) spin-only ion.

The magnetic moments calculated using Rietveld refinements at different pressures and at 30 K are summarized in the Table 2. The value obtained in the HP cell at 0 GPa, 4.84(3) μB, agrees with the value we obtained outside the HP cell at the same conditions, 4.75(3) μB. The magnetic moment at 7.7(4) GPa, 4.91(6) μB, agrees within one standard deviation with the value at 0 GPa, indicating that the magnitude of the magnetic moment is not affected much by the change of the Fe···Fe distance from 3.158(15) Å to 3.135(2) Å and the increase of the Fe···Fe···Fe angle from 101.99(7)° to 103.87(2)°. However, comparing 0 GPa with 7.74GPa the magnetic moment presents a difference of 4.3(4)°, with the magnetic moments pointing along *a* direction and remaining contained in the *ac*-plane.

| Pressure (GPa) | m(μB) | Phi(°) | Theta(°) |
|---|---|---|---|
| 0 | -4.84(3) | 0 | 61.7(3) |
| 3.9(4)† | -5.03(3) | 0 | 61.7(2) |
| 6.3(4)† | -4.96(7) | 0 | 57.3(3) |
| 7.7(4)* | -4.91(6) | 0 | 66.0(4) |
| 8.7(4)† | -5.69(6) | 0 | 55.7(3) |

**Table 2.** Magnetic moments of $FeWO_4$ obtained at 30 K and different pressure by the Rietveld refinement using spherical coordinates. *Measurements in the second load. †The data obtained for the pressures suffered from dark areas due to a partial rupture of the cell.

At the other three pressures, we obtained magnetic moments which are slightly larger than the value obtained at 0 GPa or 7.7 GPa (second load), we consider that these values are affected by experimental artifacts. The diffraction pattern obtained for these pressures suffered from dark areas, which produce an incorrect estimation of some observed reflections that gives rise to inaccurate models. The deformations in the gaskets cause the displacement of the cadmium plates used to eliminate the scattering from the pressure cell. This displacement produces partial shielding of the neutrons scattered by the sample, but only in certain directions, preventing an accurate Rietveld refinement. The fact that our experiments indicate that the magnetic moment remains unaffected by a pressure of 7.74 GPa suggests that to fully understand the magnetism of FeWO$_4$ under high pressure, one must consider a significantly more complex model than the one currently used to describe its magnetic properties at ambient pressure.

Through these studies we have also determined the Néel temperature at different pressures. Thermogram at 7.74 GPa is shown in Figure 7 together with the pressure dependence of T$_N$. From the measurements carried out, we can conclude that T$_N$ is affected by pressure, in the range studied in the present work a temperature shift of 5(1) K is observed, see Figure 7. Our results also suggest that the behavior of the Néel temperature in FeWO$_4$ under HP shows a similar dependence than that observed for the AF$_1$ phase in MnWO$_4$ [27]. The Néel temperature of the AF$_1$ phase increases from ambient pressure to 2 GPa at a rate of 1K/GPa [27], in our case the temperature shift has the same tendency but with a slightly lower slope. The effect of the external pressure in FeWO$_4$ produces slight distortions in the lattices affecting the interatomic distances and the magnetic exchange interaction parameters, although the magnetic phase diagram of FeWO$_4$ is less complex than that of MnWO$_4$, the tendency in the corresponding phase (AF$_1$) is very similar.

To conclude, we would like to comment on the effect of pressure in the crystal structure of FeWO$_4$ at 300 K. Despite having few data points, we have proceeded to determine an EoS using a Birch-Murnaghan model of the third order [28] given the behavior obtained through the analysis of the normalized stress (F) versus the finite strain (f) [29], which is shown in Figure 8. The corresponding fit is shown in Figure 9 compared with results from pervious X-ray diffraction (XRD) experiments and density functional theory calculations [14]. The figure shows that the present data shows a slightly smaller decrease of the volume with pressure than observed in previous XRD experiments [14]. In fact, the present neutron results align better with density-functional theory simulations than previous XRD experiments [14]. Given the limited number of data points measured in the present experiments, in the fit of the EoS we fixed the value of the pressure derivative of the pressure derivative of the bulk modulus to $K_0^{'} = 5$, taking this value from the previous XRD study. From the fit to the present results, we obtained a value of the bulk modulus of $K_0 = 162(6)\ GPa$.

This value is closer to the value obtained from computing simulations, $K_0$ = 150(1) $GPa$, than to the value determined from XRD experiments 136(2) $GPa$. The difference between the bulk modulus reported from XRD experiments and the present study is 16%. It can be caused be several reasons, the use of a different PTM, the use of a synthetic sample in this study and a natural mineral in XRD studies, and the different pressure range measured in both experiments, 20 GPa in XRD and 8.7(4) GPa in the present study. High-pressure XRD studies using He as pressure medium [30] are needed for a more accurate determination of the EoS of $FeWO_4$.

## CONCLUSIONS

We have reported the first high-pressure and low-temperature neutron diffraction study of $FeWO_4$ up to 8.7(4) GPa and 30.0(5) K. The experiments were performed on a Paris–Edinburg large volume press on the XtremeD diffractometer at the Institut Laue Langevin. The magnetic propagation vector, **k** = (1/2, 0, 0), of $FeWO_4$ remains invariant from ambient pressure to 8.7GPa. The magnetic structure could be described in the $P_a2/c$ Shubnikov space groups with the magnetic moments contained in the *ac*-plane. As a function of the applied pressure, the magnetic moments undergo a shitf of 4.3(4)º resulting in a magnetic structure where the moments are deviated by 28.3(3)º with respect to the *a*-axis. However, the magnitude of the magnetic moments doesn't show a significant variance. The ordering temperature, $T_N$, is affected by pressure, increasing by about 5(1)K in the pressure range studied. This increase in the $T_N$ exhibits a behavior akin to that previously in the $MnWO_4$. The origin of this behavior is the modification of the exchange coupling due to the reduction of the Fe⋯Fe distance and the increase of the Fe⋯Fe⋯Fe angle. In addition, the octahedron distortion due to the Jahn-Teller distortion produces a non-negligible orbital contribution that is the responsible for final magnetic moment, which is slightly larger than the expected for an S = 2 ion with spin-only contribution. As observed in the $MnWO_4$ compound, the pressure stabilizes the commensurate phase ($AF_1$), therefore the range in which the incommensurate phase ($AF_2$) is observed decreases with increasing pressure. In the case of $FeWO_4$ compound, there are no incommensurate phase, but increasing the pressure is not expected to be a good strategy to obtain this frustrated phase.

Alternatively, local or anisotropic distortions could be used, via solid solutions or non-hydrostatic pressure/deviation conditions. Finally, we determined the pressure-volume equation of state at 300 K and compared it with previous results obtained using different methods.

# ACKNOWLEDGEMENTS

The authors thank D. Vie and R. Vilaplana for synthesizing the sample used for the studies. The authors thank the ILL for providing access to the XtremeD instrument through proposals 5-31-3031 (10.5291/ILL-DATA.CRG-3106) and 5-15-639 (10.5291/ILL-DATA.5-15-639). They also thanks to Dr. Clemens Ritter for providing access to the D1B instrument. The also thanks the technical assistance of the scientific and technical staff of ILL. D.E. acknowledges the financial support from the Spanish Research Agency (AEI) and Spanish Ministry of Science, Innovation, and Universities (MCIU) under Project PID2022-138076NB-C41 and RED2022-134388-T (DOI: 10.13039/501100011033). D.E. also thanks the financial support of the Generalitat Valenciana (GVA) under grants PROMETEO CIPROM/2021/075 (GREENMAT) and MFA/2022/007. This study forms part of the Advanced Materials program and is supported by MCIU with funding from European Union Next Generation EU (PRTR-C17.I1) and by GVA.

# AUTHOR DECLARATIONS

**Conflict of Interest**

The authors have no conflicts to disclose.

**Author Contributions**

Oscar Fabelo: Investigation (equal); Methodology (equal); Writing – original draft (equal). Javier Gonzalez-Platas: Investigation (equal); Methodology (equal); Writing – review & editing (equal). Stanislav Savvin: Investigation (equal); Methodology (equal). Pablo Botella: Investigation (equal); Methodology (equal). Daniel Errandonea: Investigation (equal); Writing – review & editing (equal).

## DATA AVAILABILITY

The data that support the findings are deposited in the Institut Laue-Langevin (ILL) and are openly available in doi:10.5291/ILL-DATA.CRG-3106 and doi:10.5291/ILL-DATA.5-15-639.

Figures

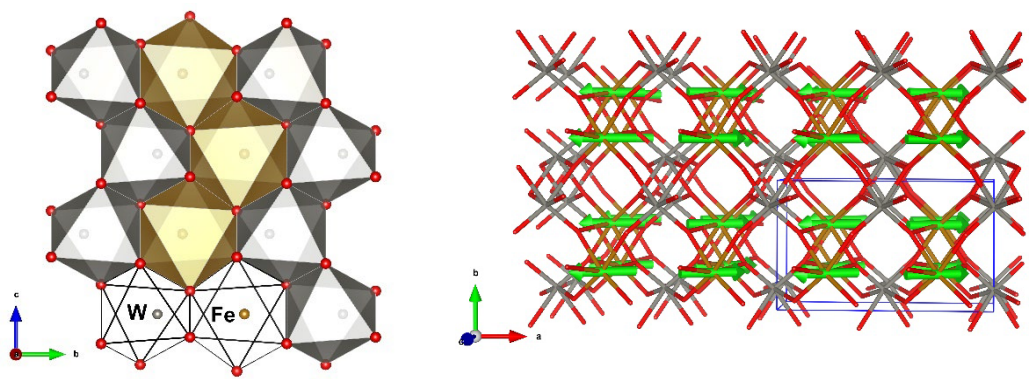

**Figure 1.** (Left) Crystal structure of FeWO$_4$ at room temperature. Fe, W, and O atoms are shown in gold, grey, and red color, respectively. The FeO$_6$ and WO$_6$ octahedra are represented. (Right) Magnetic structure of FeWO$_4$ at low temperature. The magnetic unit cell has been highlighted in blue.

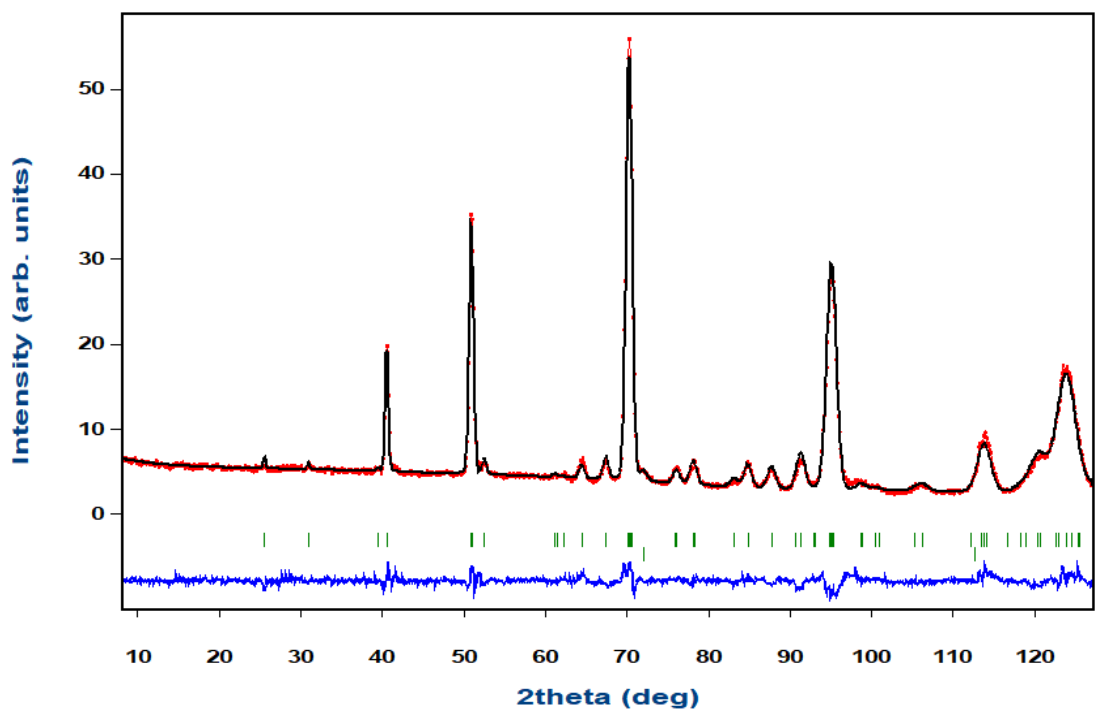

**Figure 2.** Rietveld refinement of FeWO$_4$ under room-temperature conditions, measured on the D1B diffractometer. Experimental data and fitted profiles are represented by red and black lines, respectively, while the blue line indicates the residuals of the refinement. Vertical green ticks mark the positions of the Bragg reflections. The second phase corresponds to V peaks originating from the sample holder, which were refined using the LeBail method. The final Bragg R-factor obtained for the paramagnetic FeWO$_4$ phase in the last refinement cycle was 6.00%.

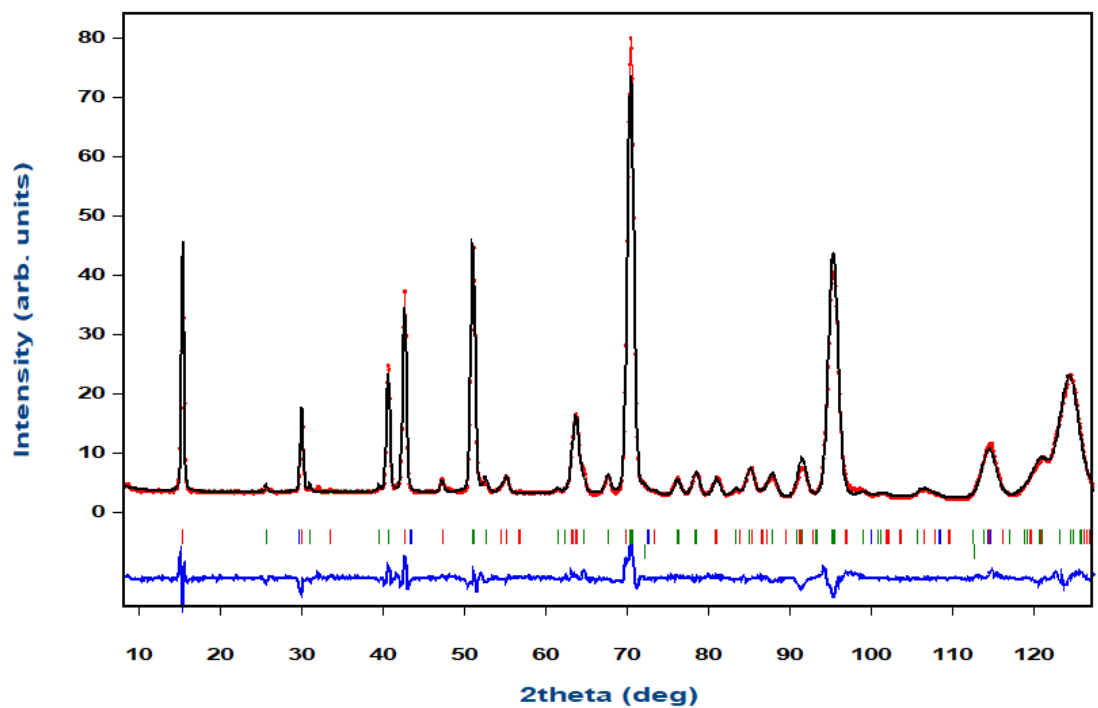

**Figure 3.** Rietveld refinement of FeWO$_4$ at 30K under ambient pressure, utilizing the $\Gamma_2$ irreducible representation corresponding to the magnetic space group $P_a2/c$. The second phase corresponds to the V peaks originating from the sample holder, which were refined using the Le Bail method. The obtained Bragg R-factors of the FeWO$_4$ phase was 3.45%, including nuclear and magnetic structure. Experimental data and fitted profiles are represented by red and black lines, respectively, with the blue line indicating the residual of the refinement. Green vertical ticks indicate the structural Bragg peaks, while red vertical ticks mark the positions of the magnetic Bragg peaks.

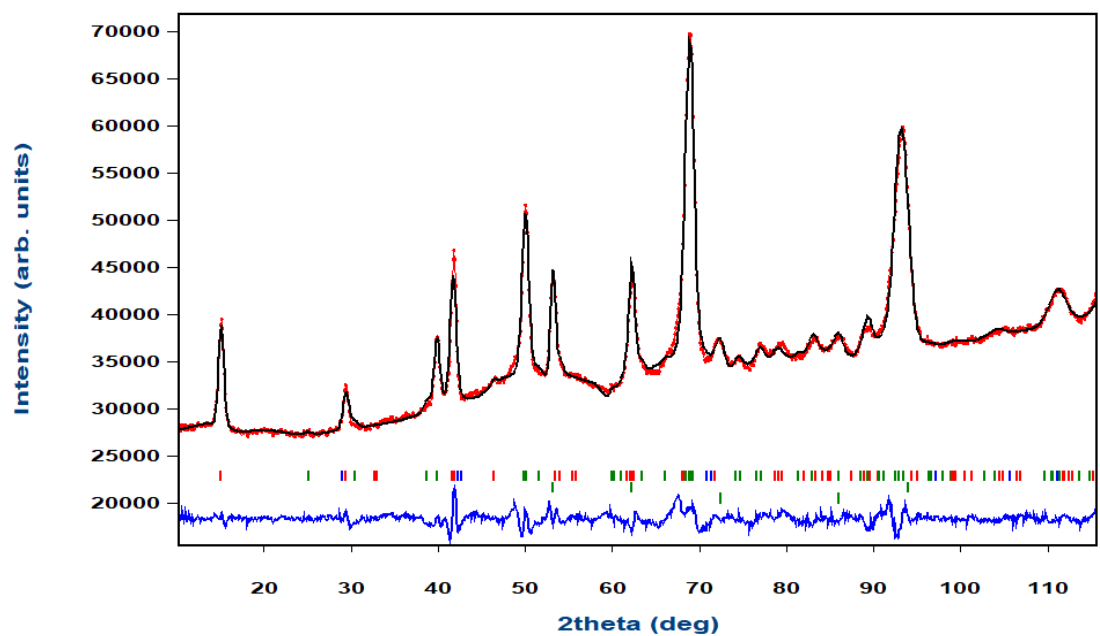

**Figure 4.** Rietveld refinement of $FeWO_4$ at 30 K and 7.7 GPa, employing the same $\Gamma_2$ irrep. Rietveld refinement was applied for both $FeWO_4$ and cBN, while the lead phase was fitted using the Le Bail method. The obtained Bragg R-factors are 3.45% for $FeWO_4$, 2.58% for cBN, and 0.36% for Pb. Experimental data and fitted profiles are represented by red and black lines, respectively, with the blue line indicating the residual of the refinement. Green vertical ticks indicate the structural Bragg peaks, while red vertical ticks mark the positions of the magnetic Bragg peaks. The second and third sets of green vertical ticks in the lower part of the pattern correspond to the lead and cBN phases, respectively.

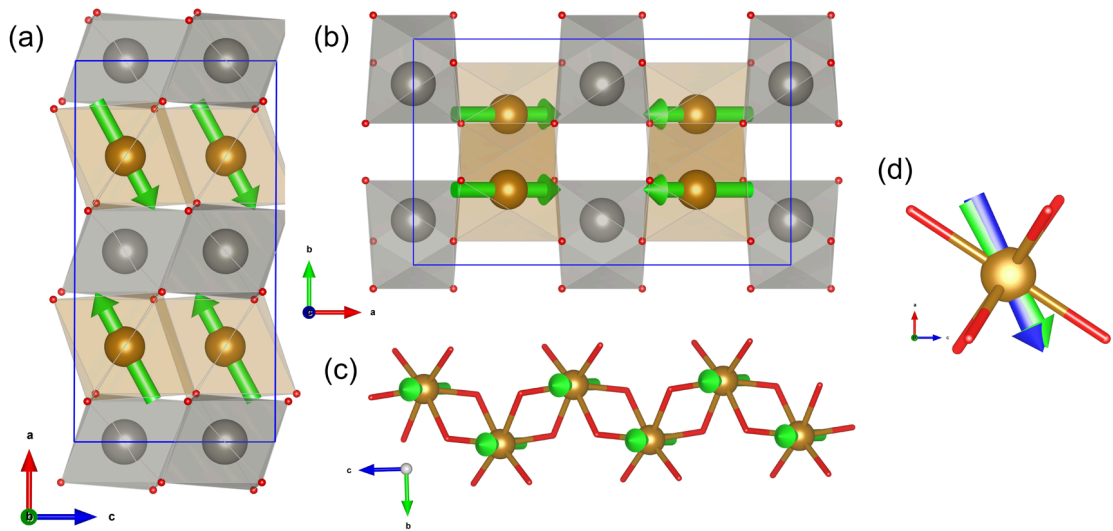

**Figure 5.** (a) View along the *b*-axis and (b) view along the *c*-axis of the proposed magnetic model for $FeWO_4$ at 30 K under ambient pressure. (c) A detailed representation of the zig-zag iron chain extending along the *c*-axis, where all magnetic moments are ferromagnetically coupled through a double *µ*-oxo bridge. (d) A comparison of the magnetic moment orientations at 30 K and ambient pressure (green arrow) with those obtained at 30 K under 7.7 GPa (blue arrow). Fe, W, and O atoms are shown in gold, grey, and red color, respectively. In figure (a) and (b) the W-O and Fe-O bonds have been omitted for the sake of clarity.

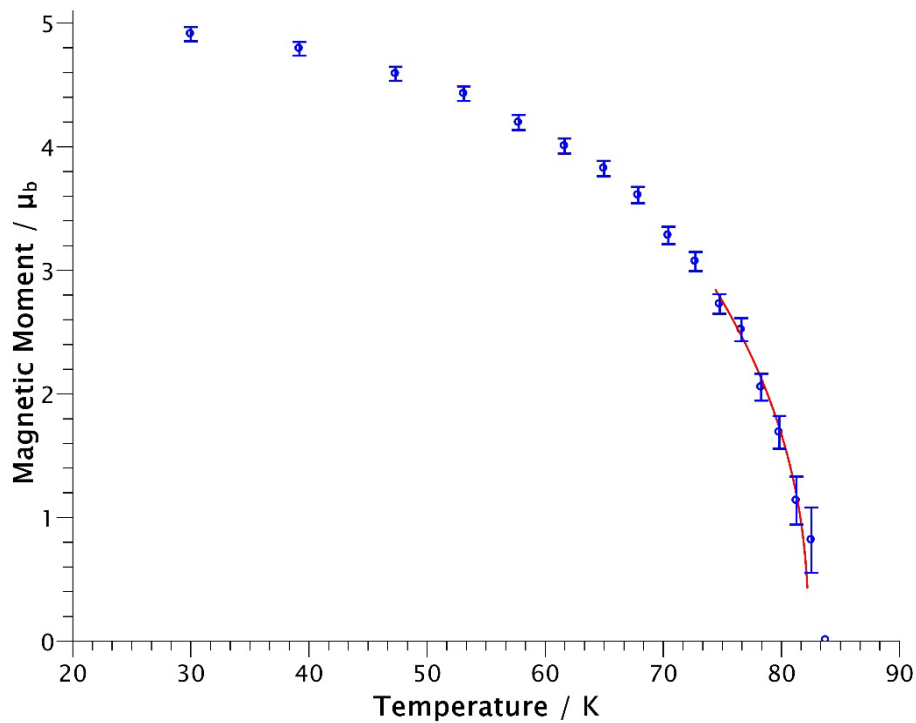

**Figure 6.** Evolution of the magnetic moment as a function of the temperature. The solid line represents the power-law fit of the data in the critical region (see main text).

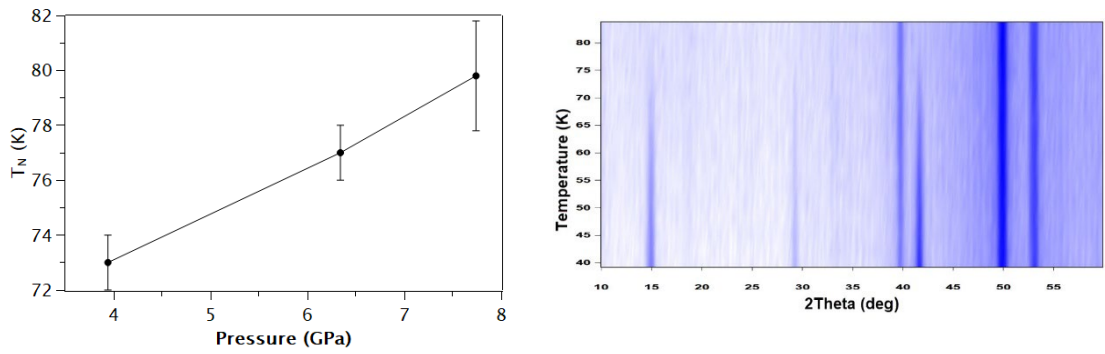

**Figure 7.** (Left) Values obtained for $T_N$ (K) during measurements by lowering the temperature. (Right) Thermogram at 7.7(4) GPa.

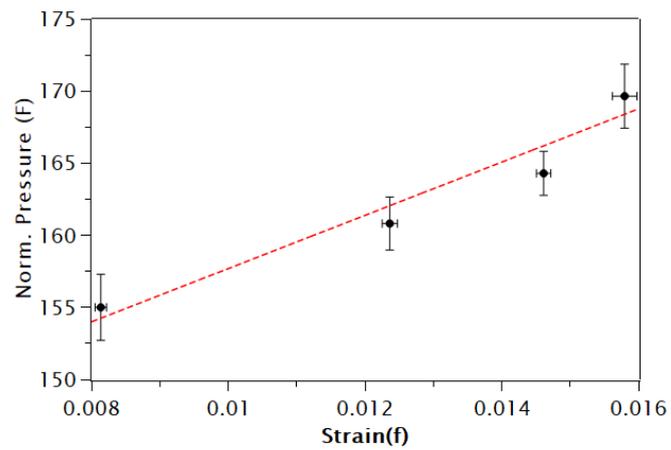

**Figure 8.** f-F plot. Symbols are the results from experiments and the red dashed line the fit.

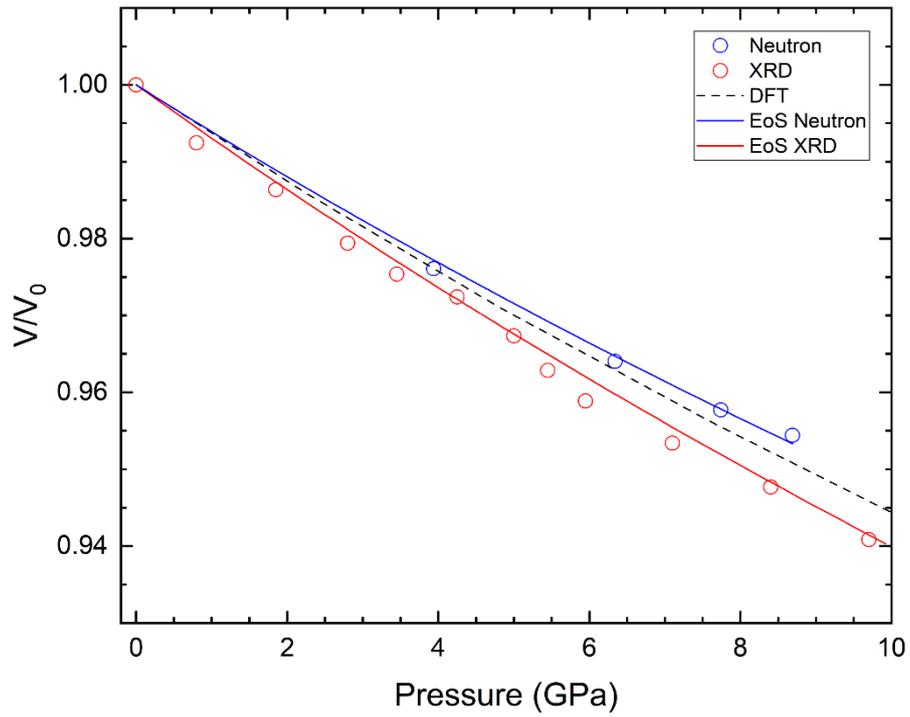

**Figure 9.** Comparison of the pressure dependence of the volume determined from present neutron diffraction experiments (blue circles), XRD experiments (red circles), and DFT calculations (black dashed line). The blue and red solid lines are the EoS determined from neutron and XRD experiments, respectively, which are described in the text.